\documentclass[lettersize,journal]{IEEEtran}
\usepackage{amsmath,amsfonts}
\usepackage{algorithmic}
\usepackage{algorithm}
\usepackage{array}
\usepackage[caption=false,font=normalsize,labelfont=sf,textfont=sf]{subfig}
\usepackage{textcomp}
\usepackage{stfloats}
\usepackage{booktabs}
\usepackage{adjustbox}
\usepackage{subcaption}
\usepackage{tabularx}
\usepackage{multicol}
\usepackage{multirow}
\usepackage{url}
\usepackage{threeparttable}
\usepackage{verbatim}
\usepackage{graphicx}
\usepackage{cite}
\usepackage{xurl}
\usepackage{hyperref}
\hypersetup{
 colorlinks=true,
 linkcolor=black,
 filecolor=black, 
 urlcolor=black,
 citecolor=black
 }

\usepackage{enumitem,amssymb}
\newlist{todolist}{itemize}{2}
\setlist[todolist]{label=$\square$}
\usepackage{pifont}

\usepackage{import}
\usepackage{longtable}
\usepackage{amsmath}
\usepackage[noabbrev]{cleveref}
\begin{document}

\title{Architectural Design Decisions for Data Ingestion in Big Data Architectures}

\author{Yikang Huang, Stefano Fossati, Indika Kumara, Damian A. Tamburri, and Willem-Jan van den Heuvel%
\thanks{Yikang Huang, Stefano Fossati, Indika Kumara, Damian A. Tamburri, and Willem-Jan van den Heuvel are with the Jheronimus Academy of Data Science (JADS), 's-Hertogenbosch 5211 DA, The Netherlands.}%
\thanks{Stefano Fossati and Willem-Jan van den Heuvel are with Eindhoven University of Technology (TU/e), Eindhoven 5600 MB, The Netherlands.}%
\thanks{Yikang Huang, Indika Kumara, and Willem-Jan van den Heuvel are with Tilburg University, Tilburg, The Netherlands.}%
\thanks{Damian A. Tamburri is with University of Sannio, Benevento 82100, Italy.}%
\thanks{Contacts: Stefano Fossati (s.fossati@tue.nl) and Indika Kumara (i.p.k.weerasinghadewage@uvt.nl).}
}

% The paper headers
\markboth{This is a preprint under review. Manuscript submitted to IEEE Transactions}%
{Huang \MakeLowercase{\textit{et al.}}}

% \IEEEpubid{0000--0000/00\$00.00~\copyright~2021 IEEE}
% Remember, if you use this you must call \IEEEpubidadjcol in the second
% column for its text to clear the IEEEpubid mark.

\maketitle

\begin{abstract}
\noindent\textbf{Introduction:} Data ingestion is the first critical step in a data architecture, in which data from various sources is moved to storage systems, such as data lakes and data warehouses, for analysis and insight generation. When developing a data ingestion pipeline, practitioners (e.g., data engineers and architects) must understand the data ingestion requirements (functional and non-functional) and select appropriate design options to support them. 

\noindent\textbf{Objective:} To help practitioners systematically develop data ingestion processes, this paper aims to compile a catalog of data ingestion design decisions, their options, and the selection drivers. 

\noindent\textbf{Method:} We conducted a systematic gray literature review on data ingestion, analyzed the selected literature using qualitative data analysis methods, and conducted semi-structured interviews with experts to assess the framework's practicality and utility.

\noindent\textbf{Results:} From 42 gray literature articles, we identified 8 architectural design decisions, 59 decision options, and 36 decision drivers to guide practitioners in the design and implementation of data ingestion pipelines. The framework was evaluated through expert interviews, assessing its completeness, simplicity, understandability, operational feasibility, and usefulness.

\end{abstract}

\begin{IEEEkeywords}
Data ingestion , Data architecture , Big data , Design decisions , Gray literature
\end{IEEEkeywords}

\section{Introduction}
\label{sec:introduction}
Data ingestion collects and moves raw data from various sources into target data processing environments, ensuring the data is readily available for implementing data use cases in organizations~\cite{reis2022fundamentals}. It is the initial component and process in a big data pipeline~\cite{davoudian2020big}. Architectural deficiencies in the data ingestion process can result in low-quality data being fed into the analytics pipelines~\cite{ipekdataquality}. For example, the data may be ingested only partially (data loss), multiple times (data duplicates), or incorrectly formatted (data inaccuracy)~\cite{ipekdataquality}. A change to a data source's schema can break downstream data pipelines without the necessary provisions in the data architecture to detect and address schema evolution~\cite{rucco2025mind}. Due to insufficient monitoring and governance of data ingestion processes, a data consumer may gain access to data containing sensitive attributes, thereby violating data privacy regulations~\cite{9113446}.     

A catalog of relevant Architectural Design Decisions (ADDs) and their options can help data engineers design and implement efficient data ingestion solutions that meet specific functional and non-functional requirements, while preventing issues such as those mentioned above.  Prior research on data ingestion has investigated topics such as developing tailored ingestion solutions, addressing schema evolution and data heterogeneity~\cite{rucco2025mind,marcu2018kera,9113446,sreepathy2024data}, assessing data ingestion quality~\cite{ipekdataquality}, and data ingestion in specific environments~\cite{singh2026llm}. However, none provide a comprehensive catalog of design decisions and their implementation options for data ingestion. On the other hand, existing studies on ADD frameworks~\cite{warnett2022architecturalworkflow, Zhang2022ArchitectureDI,10592671} only mention data ingestion as a component, with little elaboration on its ADDs. To address this critical research gap, this study aims to answer the following research questions:
\begin{itemize}
    \item \textbf{RQ$_1$} What ADDs, decision options, and decision drivers are present in the context of the data ingestion components in big data architectures, and how are they related?
    \item \textbf{RQ$_2$} To what extent do data engineering experts agree with the identified ADDs, decision options, and decision drivers?
\end{itemize}

To investigate these questions, we first conducted a systematic review of gray literature~\cite{garousi2019guidelines} on data ingestion, identifying 42 articles out of 115 sources.  Next, we analyzed the selected articles using qualitative methods to identify ADDs, decision options, and drivers that influence the design of the data ingestion component in a data architecture. We compiled a catalog of 8 distinct ADDs, 59 decision options, and 36 decision drivers. Our framework provides practitioners with guidance for designing an effective data ingestion solution that incorporates its best practices. To manage the effort required to analyze the literature, we decided only to focus on gray literature. A preliminary analysis of academic and gray literature on data ingestion showed that gray literature can provide rich insights into the various decisions practitioners make when designing data ingestion processes within big data architectures. Moreover, we observed that many existing ADD frameworks rely on gray literature~\cite{warnett2022architecturalworkflow, Zhang2022ArchitectureDI,10592671}.  

Secondly, to evaluate the framework's practical applicability and usefulness, we conducted semi-structured interviews with five experts with extensive experience in related fields. The interviews were designed to assess five key evaluation criteria: \emph{completeness, simplicity, understandability, operational feasibility, and usefulness}. The collected interview data were systematically analyzed using thematic analysis, with the identified themes compared and aligned with the findings from the gray literature review. The results of this validation process provided evidence supporting the relevance, applicability, and practical value of the proposed architectural design guidance framework for data ingestion.

\textbf{Paper Structure} \Cref{sec:related} reviews the related studies. \Cref{sec:research-design} describes our research design. \Cref{sec:framework} presents our ADD framework. \Cref{sec:interview} presents the insights from the semi-structured expert interview. \Cref{sec:discussion} discusses the threats to the validity. \Cref{sec:conc} concludes the paper. 

\section{Related Work}
\label{sec:related}

We consider prior work on architectural solutions for data ingestion and on ADD frameworks in general. 

Rucco et al.~\cite{rucco2025mind} proposed a design pattern for flexible data ingestion, in which metadata about data sources and ingestion pipelines (e.g., data schemas and data transformation rules) is used to dynamically generate and execute the ingestion code for a given data source.  Marcu et al.~\cite{marcu2018kera} applied the strategy of dynamically partitioning a data stream into fixed-sized batches to adapt data ingestion to the changing numbers of data producers and consumers. Ranchal et al.~\cite{9113446} developed a data ingestion service to support the migration of healthcare data to the cloud. They identified the functional and quality requirements specific to healthcare data ingestion and designed a layered architecture to support them. The layers are data acquisition, data parsing and validation, and data storage and auditing. Sreepathy et al.~\cite{sreepathy2024data} proposed a data ingestion-as-a-service for a data lake that comprised three layers: data source connectivity, configurable data transformation logic, and metadata management. Ustunboyacioglu et al.~\cite{ipekdataquality} proposed an extension to the data lake architecture to support separate quality assessments of data at ingestion and after data processing. Their architecture utilized metadata of data sources and data quality testing rules. Singh et al.~\cite{singh2026llm} employed large language models (LLMs) to semi-automatically resolve data schema mismatches when ingesting data from different organizations into shared data spaces.

Several ADD frameworks exist for machine learning systems and data architectures. Zhang et al.~\cite{Zhang2022ArchitectureDI} presented an empirical study of architectural decisions in AI (Artificial Intelligence)-enabled systems using Stack Overflow data. They identified 13 general decisions, including data, deployment, and integration.  Warnett and Zdun~\cite{warnett2022architecturaldeployment,warnett2022architecturalworkflow} used gray literature on machine learning (ML) model deployment and pipelines to identify the related ADDs, decision options, and drivers. Data ingestion was treated as a single ADD with four decision options, focusing on the data ingestion method (batch vs. stream).  Eijk van et al.~\cite{10592671} also employed gray literature to develop an ADD framework for data platforms within data mesh architectures. Data ingestion was identified as a feature that the platform should support in a self-service manner. 

Overall, the importance and challenges of designing data ingestion solutions to meet specific functional and non-functional goals have been acknowledged in the current literature. While some ADD frameworks include data ingestion as a decision, there are no studies that systematically enumerate a comprehensive set of ADDs for data ingestion.  This study aims to address the research gap.

\section{Research Design}
\label{sec:research-design}

In this study, we first conducted a systematic gray literature review (SGLR) on data ingestion in big data architectures. Secondly, we analyzed the selected sources to extract ADDs, along with their associated options and forces, following established approaches~\cite{warnett2022architecturaldeployment,warnett2022architecturalworkflow,10592671}. And finally, we conducted semi-structured interview study with practitioners to validate the results extracted from the gray literature review.

\smallskip
\noindent\textbf{Design Framework Creation.} The SGLR followed the guidelines of Garousi et al.~\cite{garousi2019guidelines} and prior SGLR studies~\cite{goedegebuure2024data,kumara2021s}. The search was conducted using Google in Incognito mode, complemented by additional relevant literature from backlinks and reference pages~\cite{garousi2019guidelines}. Inclusion and exclusion criteria were applied to assess source quality and refine the initial literature corpus. We selected these criteria from the existing reviews~\cite{kumara2021s,goedegebuure2024data,garousi2019guidelines}. We included sources written in English, fully accessible, and aligned with the focus of this study (e.g., data ingestion decisions and drivers). Gray sources that did not meet the quality assessment criteria  (e.g., based on the publisher's reputation, the author's expertise and employment organization, and the depth of the content) were excluded. Regarding the search query, we first tried various keywords, including data ingestion, data collection, data gathering, and data movement. However, except for \textit{data ingestion}, other keywords returned many irrelevant results. Hence, we decided to use only the keyword \textit{data ingestion}, as it is widely used by both practitioners and researchers.

The first and the second author independently applied the inclusion/exclusion criteria, and after the second iteration, the calculated Cohen’s Kappa coefficient~\cite{cohen1960coefficient} was 0.6817, indicating substantial agreement. Following conflict resolution, the 115 sources were reduced to 62 based on quality assessment criteria, and further to 42 after post-quality assessment. To identify ADDs, options, and drivers, the selected sources were analyzed through an iterative coding process~\cite{wieringa2014empirical} using the qualitative analysis software Atlas.ti~\footnote{https://atlasti.com/}. In particular, open coding, axial coding, and selective coding were performed until theoretical saturation was reached based on grounded theory~\cite{glaser2017discovery,corbin1990grounded}.

\begin{table*}[t]
\scriptsize
\centering
\begin{tabular}{@{}lll@{}}
\toprule
\textbf{ID} & \textbf{Source Type}      & \textbf{Title}                                                                                                                                  \\ \midrule
\href{https://www.techtarget.com/whatis/definition/data-ingestion#:~:text=Data%20ingestion%20is%20the%20process,as%20the%20source%20emits%20it.}{s1}        & Blog                   & Data ingestion   \\
\href{https://www.striim.com/blog/what-is-data-ingestion-and-why-this-technology-matters/}{s2}        & Blog                   & What is Data Ingestion and Why This Technology Matters   \\
\href{https://streamsets.com/learn/data-ingestion/}{s3}        & Article                   & Data Ingestion: Tools, Types, and Key Concepts   \\
\href{https://www.informatica.com/nl/resources/articles/what-is-data-ingestion.html}{s4}        & Article                   & What Is Data Ingestion? Learn Key Use Cases, Capabilities, and Tools   \\
\href{https://docs.aws.amazon.com/whitepapers/latest/building-data-lakes/data-ingestion-methods.html}{s5}        & White Paper                   & Data ingestion methods   \\
\href{https://learn.microsoft.com/en-us/azure/data-explorer/ingest-data-overview}{s6}        & Doc                   & Azure Data Explorer data ingestion overview   \\
\href{https://www.projectpro.io/article/data-ingestion/600}{s7}        & Article                   & Data Ingestion-The Key to a Successful Data Engineering Project   \\
\href{https://hevodata.com/learn/data-ingestion-pipeline/}{s8}        & Article                   & Data Ingestion Pipeline: Comprehensive Guide 101   \\
\href{https://hevodata.com/learn/data-ingestion-methods-for-data-lake/}{s9}        & Article                   & Ultimate Guide on the Best Data Ingestion Methods for Data Lakes   \\
\href{https://scaleyourapp.com/what-is-data-ingestion-how-to-pick-the-right-data-ingestion-tool/}{s10}        & Article                   & What is Data Ingestion? How to Pick the Right Data Ingestion Tool?   \\
\href{https://www.snowflake.com/blog/best-practices-for-data-ingestion/}{s11}        & Article                   & Best Practices for Data Ingestion with Snowflake: Part 1   \\
\href{https://atlan.com/data-ingestion-vs-data-integration/}{s12}        & Article                   & Data Ingestion vs. Data Integration: How Are They Different?   \\
\href{https://airbyte.com/blog/best-practices-data-ingestion-pipeline}{s13}        & Blog                   & Best Practices to Design a Data Ingestion Pipeline   \\
\href{https://www.kai-waehner.de/blog/2022/07/05/data-streaming-for-data-ingestion-into-data-warehouse-and-data-lake/}{s14}        & Blog                   & Data Streaming for Data Ingestion into the Data Warehouse and Data Lake   \\
\href{https://databand.ai/blog/data-ingestion-strategy/}{s15}        & Blog                   & Your Data Ingestion Strategy Is A Key Factor In Data Quality   \\
\href{https://www.analyticssteps.com/blogs/what-data-ingestion-challenges-and-types}{s16}        & Blog                   & What is Data Ingestion? Challenges and Types   \\
\href{https://restapp.io/blog/data-ingestion-meaning/}{s17}        & Blog                   & What is Data Ingestion? Meaning, Benefits and Best Practices   \\
\href{https://coefficient.io/top-11-data-ingestion-tools-to-jumpstart-your-data-strategy/}{s18}        & Blog                   & Top 11 Data Ingestion Tools to Jumpstart your Data Strategy   \\
\href{https://neptune.ai/blog/data-ingestion-and-feature-store-component-mlops-stack}{s19}        & Blog                   & How to Solve the Data Ingestion and Feature Store Component of the MLOps Stack   \\
\href{https://itnext.io/big-data-ingestion-options-2d95851410d3}{s20}        & Blog                   & Big Data Ingestion Options   \\
\href{https://www.ascend.io/data-ingestion/}{s21}        & Article                   & What Is Data Ingestion?   \\
\href{https://www.indeed.com/career-advice/career-development/data-ingestion-tools}{s22}        & Article                   & 19 Data Ingestion Tools (Plus Benefits and Features)   \\
\href{https://ezdatamunch.com/what-is-data-ingestion/}{s23}        & Blog                   & What Is Data Ingestion? Know the benefits of data ingestion   \\
\href{https://azure-training.com/azure-data-science/overview-of-data-ingestion-in-azure-data-explorer/}{s24}        & Blog                   & Overview of data ingestion in Azure Data Explorer   \\
\href{https://insights.daffodilsw.com/blog/what-is-data-ingestion-and-why-is-it-essential}{s25}        & Blog                   & What Is Data Ingestion And Why Is It Essential?   \\
\href{https://hightouch.com/blog/what-is-data-ingestion-the-definitive-guide#data-ingestion-tools}{s26}        & Blog                   & What is Data Ingestion? | The Definitive Guide   \\
\href{https://www.matillion.com/resources/blog/an-introduction-to-data-ingestion}{s27}        & Blog                   & An Introduction to Data Ingestion   \\
\href{https://druid.apache.org/docs/latest/ingestion/index.html}{s28}        & Manual                   & Ingestion   \\
\href{https://improvado.io/blog/top-data-ingestion-tools}{s29}        & Blog                   & Top 16 Data Ingestion Tools to Jumpstart your Data Strategy   \\
\href{https://www.linkedin.com/pulse/realtime-data-ingestion-an-overview-sujithkumar-chandrasekaran?trk=pulse-article}{s30}        & Blog                   & Realtime data ingestion-an overview   \\
\href{https://dataintegrationinfo.com/what-is-data-ingestion/}{s31}        & Blog                   & What is Data: Process, Tools, and Challenges Discussed   \\
\href{https://www.analyticsvidhya.com/blog/2022/06/data-ingestion-featuring-aws/}{s32}        & Article                   & Data Ingestion Featuring AWS   \\
\href{https://blog.hubspot.com/marketing/data-ingestion}{s33}        & Blog                   & Data Ingestion: What It Is Plus How And Why Your Business Should Leverage It   \\
\href{https://www.upsolver.com/blog/7-guidelines-ingesting-big-data-lakes}{s34}        & Blog                   & Data Lake Ingestion: 7 Best Practices   \\
\href{https://whatagraph.com/blog/articles/data-ingestion}{s35}        & Blog                   & What is Data Ingestion?   \\
\href{https://engineering.grab.com/real-time-data-ingestion}{s36}        & Article                   & Real-time data ingestion in Grab   \\
\href{https://www.dropbase.io/post/7-tools-to-automate-your-data-ingestion}{s37}        & Post                   & 7 Tools to Automate your Data Ingestion   \\
\href{https://docs.aws.amazon.com/whitepapers/latest/aws-cloud-data-ingestion-patterns-practices/data-ingestion-patterns.html}{s38}        & White Paper                   & AWS Cloud Data Ingestion Patterns and Practices   \\
\href{https://www.ericsson.com/en/reports-and-papers/ericsson-technology-review/articles/data-ingestion-architecture-for-telecom}{s39}        & Article                   & Data ingestion architecture for telecom applications   \\
\href{https://tech.justeattakeaway.com/2019/03/29/data-ingestion-at-just-eat-part-1-the-concept/}{s40}        & Blog                   & Data Ingestion At Just Eat – Part 1: The Concept   \\
\href{https://towardsdatascience.com/change-data-capture-cdc-for-data-ingestion-ca81ff5934d2}{s41}        & Blog                   & Change Data Capture(CDC) for Data Lake Data Ingestion   \\
\href{https://medium.com/abn-amro-developer/building-a-scalable-metadata-driven-data-ingestion-framework-7c43779277d4}{s42}        & Blog                   & Building a scalable metadata-driven data ingestion framework   \\

 \bottomrule
\end{tabular}
\caption{Grey Literature Selected - Data Ingestion}
\label{tab:gl_di}
\end{table*}

\begin{table}[t]
\scriptsize
\centering
\caption{Interviewed Experts}
\label{tab:experts-dmm}

\begin{tabular}{|p{0.05\linewidth} |p{0.28\linewidth}| p{0.27\linewidth} |p{0.05\linewidth}| p{0.05\linewidth}| p{0.10\linewidth}|}
        \hline
        \textbf{ID} & \textbf{Type of stakeholder} & \textbf{Organisation} & \textbf{CC} & \textbf{\#EY} \\ \hline

         p1  &   Senior Cloud Data Architect      & Consultancy Company                      & NL    & 5 \\ \hline
         p2  &   Machine Learning Engineer      & Manufacturing Company A                    & NL    & 3 \\ \hline
         p3  &   Machine Learning Engineer      & Manufacturing Company A                     & NL    & 6 \\ \hline
         p4  &   Data Engineer      & Software Company                      & NL    & 3 \\ \hline
         p5  &   Data Engineer      & Manufacturing Company B                    & AU    & 10 \\ \hline
    \end{tabular}

\begin{tablenotes}
 \item \textbf{ID}: pseudonym used to refer to the expert
 \item \textbf{CC}: expert's country of origin in terms of ISO Alpha-2 country codes
 \item \textbf{\#EY}: number of years of experience in data engineering practices
\end{tablenotes}
\end{table}

\smallskip
\noindent\textbf{Interview-based Refinement and Validation.} To assess the quality and applicability of the architectural decision models extracted from gray literature, we performed semi-structured interviews with domain experts. The participants were selected through \textit{purposive sampling}~\cite{etikan2016comparison} based on two predefined inclusion criteria. First, candidates were required to have at least 3 years of professional experience with ML platforms, including machine learning engineering and data engineering. Second, participants needed to represent relevant stakeholder groups that could benefit from or apply the proposed framework, including ML engineers, data engineers, data scientists, data architects, and researchers.

Potential interviewees were identified through multiple channels. We gathered recommendations from practitioners within a large European organization, reached out to authors of relevant gray literature publications, and searched for additional experts through LinkedIn\footnote{\url{https://www.linkedin.com/}}. In total, eight experts were contacted, of whom five agreed to participate in the study. \Cref{tab:experts-dmm} summarizes the participating experts characteristics.

The evaluation of the proposed framework was guided by five criteria adapted from existing evaluation characteristics~\cite{castillo2016preparing}: \emph{completeness, simplicity, understandability, operational feasibility, and usefulness}. These criteria were used to investigate whether the framework provides sufficient decision coverage, remains manageable and clear for practitioners, can realistically support decision-making activities, and offers practical value to stakeholders involved in ML platform development and adoption.
  
\smallskip
\noindent\textbf{Replication Package.} The replication package\footnote{\url{https://doi.org/10.5281/zenodo.21510046}} includes the literature sources, selection and quality assessment criteria, coded literature, interview protocol, anonymized transcripts, and interview coding. 

\section{Data Ingestion Design Decisions Framework}
\label{sec:framework}
\begin{figure*}[t]
    \centering
    \includegraphics[width=0.80\linewidth]{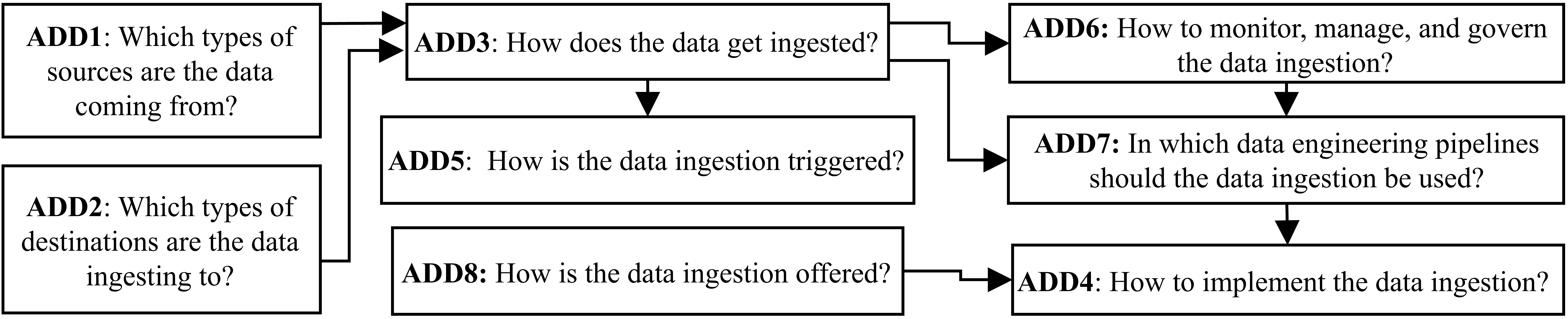}
    \caption{Relationship among data ingestion ADDs}
    \label{fig:DI_ADD_ov}
\end{figure*}

This section presents data ingestion architectural design decisions (ADDs) derived from 42 selected gray literature that are reported in Table~\ref{tab:gl_di}. We identified 8 discrete ADDs, 59 decision options, and 36 decision drivers. Figure~\ref{fig:DI_ADD_ov} shows ADDs and their relationships. \Cref{table:drivers} shows the decision selection drivers. %In the rest of this section, we discuss each ADD in detail. 

% \begin{table*}[t]
% \caption{Decision drivers}
% \label{table:drivers}
% \footnotesize

% % \tiny
% \begin{tabular}{|c|l|c|l|}
% \hline
% \textbf{Code} & \textbf{Description}& \textbf{Code} & \textbf{Description}\\ \hline
% f1& Unification and Consistency & f19 & Flexibility \\ \hline
% f2& Large Throughput and Amount of Data & f20 & Accuracy\\ \hline
% f3& Reduce Failure& f21 & Discoverability \\ \hline
% f4& Low latency & f22 & Cost Factors - Save Cost (Time and Money) \\ \hline
% f5& Reduce Complexity & f23 & Security and Privacy\\ \hline
% f6& Connectivity& f24 & Parallelism and Concurrency \\ \hline
% f7& Diversity & f25 & Ease of use \\ \hline
% f8& Functionality & f26 & Continuity\\ \hline
% f9& Resilience& f27 & Compatibility and Extendability \\ \hline
% f10 & Control & f28 & Reliability \\ \hline
% f11 & User Experience & f29 & Automation in General \\ \hline
% f12 & Cooperation and Support & f30 & Performance in General\\ \hline
% f13 & Dependency Management & f31 & Reusability \\ \hline
% f14 & Maintainability & f32 & Quality Assurance \\ \hline
% f15 & Error Handling& f33 & Risk Management \\ \hline
% f16 & Loose Coupling& f34 & Scalability, Productivity and Efficiency\\ \hline
% f17 & Deployment Flexibility& f35 & Accessibility, Availability, Fault Tolerant \\ \hline
% f18 & Recovery& f36 & Agility \\ \hline
% \end{tabular}

% \end{table*}

\begin{table*}[t]
\caption{Decision drivers}
\label{table:drivers}
\centering
\resizebox{.99\textwidth}{!}{
\begin{tabular}{|c|l|c|l|c|l|}
\hline
\textbf{Code} & \textbf{Description} &
\textbf{Code} & \textbf{Description} &
\textbf{Code} & \textbf{Description} \\ \hline

f1  & Unification and Consistency &
f13 & Dependency Management &
f25 & Ease of use \\ \hline

f2  & Large Throughput and Amount of Data &
f14 & Maintainability &
f26 & Continuity \\ \hline

f3  & Reduce Failure &
f15 & Error Handling &
f27 & Compatibility and Extendability \\ \hline

f4  & Low latency &
f16 & Loose Coupling &
f28 & Reliability \\ \hline

f5  & Reduce Complexity &
f17 & Deployment Flexibility &
f29 & Automation in General \\ \hline

f6  & Connectivity &
f18 & Recovery &
f30 & Performance in General \\ \hline

f7  & Diversity &
f19 & Flexibility &
f31 & Reusability \\ \hline

f8  & Functionality &
f20 & Accuracy &
f32 & Quality Assurance \\ \hline

f9  & Resilience &
f21 & Discoverability &
f33 & Risk Management \\ \hline

f10 & Control &
f22 & Cost Factors -- Save Cost (Time and Money) &
f34 & Scalability, Productivity and Efficiency \\ \hline

f11 & User Experience &
f23 & Security and Privacy &
f35 & Accessibility, Availability, Fault Tolerant \\ \hline

f12 & Cooperation and Support &
f24 & Parallelism and Concurrency &
f36 & Agility \\ \hline

\end{tabular}
}
\end{table*}

\subsection{ADD 1: Data Ingestion Sources}
\begin{figure*}[t]
    \centering
    \includegraphics[width=0.80\linewidth]{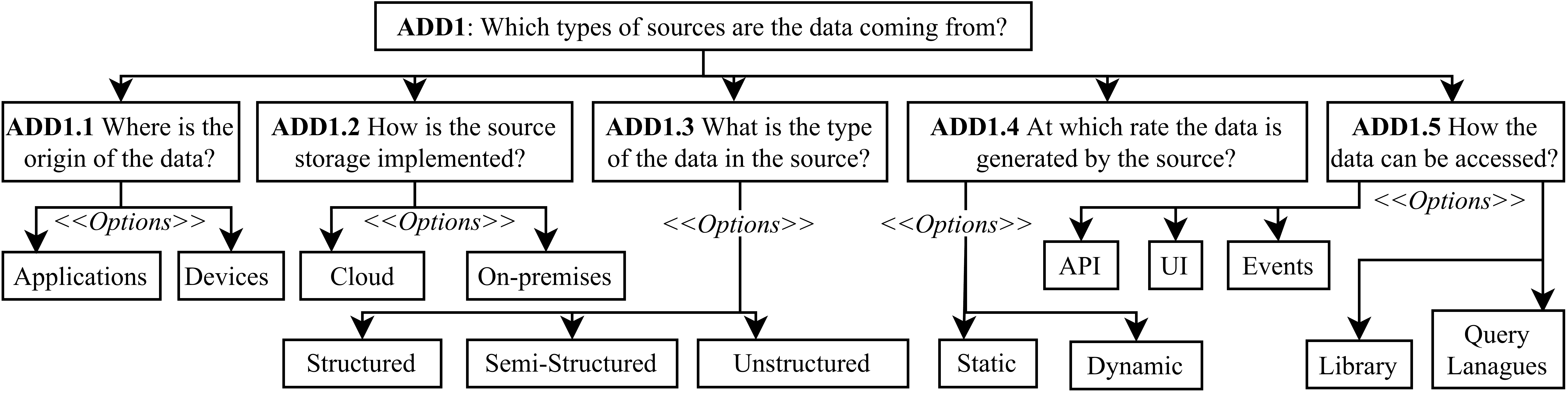}
    \caption{ADD 1: Which types of sources are the data coming from?}
    \label{fig:DI_ADD_1}
\end{figure*}

Data source systems generate data, and data ingestion processes move the generated data to destination data sinks~\cite{reis2022fundamentals}.  Figure \ref{fig:DI_ADD_1} presents the design decisions a practitioner should consider when working with source systems. Table \ref{table:DI_ADD_1} shows the evidence for each decision option (gray literature sources) and forces that can determine the selection of a given option (please refer to Table~\ref{table:drivers} for the description of codes). 

\begin{table*}[htbp]
\centering
\scriptsize
\caption{ADD 1: options, evidence, and forces}
\label{table:DI_ADD_1}
\resizebox{.99\textwidth}{!}{%
\begin{tabular}{|p{0.11\linewidth}|p{0.6\linewidth}|p{0.07\linewidth}|p{0.08\linewidth}|} 
\hline
\textbf{Options} & \textbf{Source Evidences} & \textbf{Practitioner Evidence}& \textbf{Forces}                    \\ 
\hline
Cloud-based     & s1, s2, s3, s4, s5, s7, s8, s9, s10, s11, s13, s14, s15, s16, s17, s18, s19, s21, s22,s23, s25, s26, s27, s29, s31, s32, s33, s35, s36, s37, s38, s39, s40, s41 & p4, p5 & f34,f25,f22                        \\ 
\hline
On-Premises     & s4, s5, s9, s14, s15, s16, s17, s18, s19, s22, s23, s25, s26, s27, s29, s31, s32, s33, s35, s36, s37, s38, s41  & p1, p4, p5                                               & f10,f23                            \\ 
\hline
Data Format     & s1, s2, s3, s4,s9, s10, s11, s14, s15, s16, s17,~ s18, s19, s21, s36, s37, s38, s41  & - & f2,f19                             \\ 
\hline
Applications    & s1, s2, s3, s4, s5, s6, s7, s8, s9, s10, s11, s12, s13, s14, s15, s16, s17,s18, s19, s20, s21, s22, s25, s26, s27, s29, s31, s32, s34, s35, s37, s39    & p1, p5        & f6,f7,f23                          \\ 
\hline
Devices         & s1, s2, s3, s4, s6, s7, s8, s9, s10, s12, s14, s16, s17, s19, s21, s24, s25, s26, s27, s35, s38 & p1, p2                                                                 & f2,f6.f7                           \\ 
\hline
Data Generation Rate     & s4, s6, s7, s9, s10, s11, s13, s14, s16, s18, s20, s21, s25, s26, s27, s29, s30, s32, s34, s35, s36, s40, s41                      &  -          &f2,f4,f9,f30,f36 \\ 
\hline
Access Method   & s4, s6, s7, s9, s10, s11, s13, s14, s16, s18, s20, s21, s25, s26, s27, s29, s30, s32, s34, s35, s36, s40, s41 & - & f11,f19,f29                        \\
\hline

\end{tabular}
}
\end{table*}

\textbf{The first decision is the origin of the data in source systems (ADD 1.1)}. We identified two main types: \textit{software applications} and \textit{physical devices}.  An organization may use various applications, including microservices, web applications, packaged software (e.g., ERP (Enterprise Resource Planning) tools), SaaS (Software as a Service) applications (e.g., email and ticketing systems), and external sources (e.g., social media APIs and data vendors). Regarding devices, the literature primarily discussed those used in manufacturing environments, e.g., sensors attached to production machines on the shop floor.

For an organization's internal data-source systems, developers must decide \textbf{how to implement the corresponding data storage (ADD 1.2)}, for example, for databases in microservices or ERP applications.  There are two broad options: \textit{cloud-based} and \textit{on-premises.} Note that the same decision also applies when selecting the data storage for a data sink (see the next section). A key implication is responsibility for data security, performance, and other quality issues. In the second option, organizations are solely responsible for their own data and infrastructure. According to the gray literature, the cloud-based storage option reduces upfront costs, supports massive scalability, offers diverse data storage options, and simplifies implementing complex data architectures such as data lakes, data warehouses, and data lakehouses (which combine the features of the former two in a unified architecture)~\cite{harby2022data}).  A common reason for organizations to adopt an on-premises solution is to comply with internal policies and external regulations governing the handling of sensitive data. However, organizations require data engineering experts in-house to manage complex data architectures.  

\textbf{The third decision concerns the type of data in the sources (ADD 1.3)}, with three options: \textit{structured, semi-structured, and unstructured}. Structured data has a predefined data model (e.g., an entity-relationship model) and is organized and categorized according to a well-defined schema. Relational databases are the common choice for storing it. Semi-structured data (e.g., web server logs and message payloads) supports hierarchical data modeling and metadata without enforcing a rigid schema. In contrast, unstructured data (e.g., social media feeds, images, and video streams) typically lacks a predefined data model. The reviewed literature recommends NoSQL (Not Only SQL) databases, including distributed file systems, for both semi-structured and unstructured data. 

\textbf{The fourth decision is the data generation rate at source systems (ADD 1.4)}. The literature mentions two variations:\textit{ static and dynamic rates}. The first category implies data is generated at a fixed rate, either continuously or periodically. In the second one, the sources produce data at varying rates. Data ingestion rates influence the design of ingestion methods to prevent system overload and optimize resource usage~\cite{carbone2017large}, which we will discuss in the related ADDs.  

\textbf{The fifth option is the data access methods (ADD 1.5)}. Data sources may provide multiple options: \textit{APIs, query languages, libraries, events, and web UIs}.  The first four options primarily support programmatic data access, whereas web UIs are designed for human interaction.  Cloud-based storage generally supports most of these options. For on-premises sources, data engineers may need to build and maintain custom APIs and UIs, mostly reusing existing third-party solutions. 

A data ingestion design pattern commonly mentioned by many practitioners is Change Data Capture (CDC), which enables real-time streaming of changes to source databases as events to data processing destinations.  

\subsection{ADD 2: Data Ingestion Destinations}
\begin{figure*}[t]
    \centering
    \includegraphics[width=0.8\linewidth]{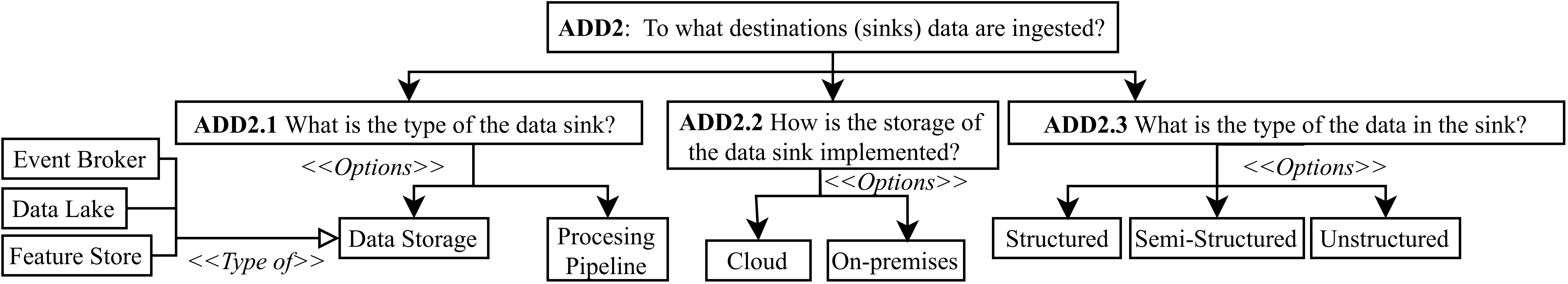}
    \caption{ADD 2: Which types of destinations are the data ingesting to?}
    \label{fig:DI_ADD_2}
\end{figure*}

The second ADD concerns the types of destinations to which data is ingested. As shown in Figure~\ref{fig:DI_ADD_2} and Table~\ref{table:DI_ADD_2}, a practitioner first needs to identify the type of data sinks for a data ingestion pipeline.  

The reviewed literature mentions two categories of \textbf{data sinks (ADD 2.1)}: \textit{data storage and data processing pipeline} (e.g., ML model training or inference pipelines, data pipelines, or feature engineering pipelines). Ingestion data storage takes various forms: a \textit{data lake} (raw or bronze data store in the data lakehouse architecture, or a staging data store in the data warehouse architecture), an\textit{ event store/broker, and a feature store}. The data lake is recommended for batch data ingestion, while the event store mainly handles data/event streams. Note that, in the data warehouse architecture, the storage of processed data can be the ultimate destination of the ingestion pipeline (see ADD 7). A feature store is a centralized repository for storing, managing, and serving ML model features. It provides a single source of truth for features, enabling data scientists and ML engineers to easily access and reuse features across different ML applications and models. 

The second and third ADDs concern the selection of \textbf{the storage for a data sink (ADD 2.2)}, especially the build-or-buy decision, and \textbf{the data format (ADD 2.3)}, which are similar to the corresponding ADDs in the data ingestion source decision discussed in the previous section.    

\begin{table*}[htbp]
\centering
\caption{ADD 2: options, evidence, and forces}
\label{table:DI_ADD_2}
\resizebox{.99\textwidth}{!}{%
\begin{tabular}{|p{0.15\linewidth}|p{0.55\linewidth}|p{0.1\linewidth}|p{0.1\linewidth}|} 
\hline
\textbf{Options}      & \textbf{Source Evidences}                                                                                                                                                              & \textbf{Practitioner Evidences} & \textbf{Forces} \\ \hline
Cloud-based Storage  & p4, s3, s4, s5, s9, s11, s14, s15, s17, s18, s19, s20, s25, s26, s27, s29, s32, s34, s36, s37, s38, s40, s41                                                                           & p4                              & f34,f25,f22     \\ \hline
On-Prem Storage      & p1, p4, s1, s2, s3, s4, s5, s6, s7, s8, s9, s11, s12, s13, s14, s15, s16, s17, s18, s19, s20, s23, s25, s26, s27, s29, s30, s31, s32, s33, s34, s35, s36, s37, s38, s39, s40, s41, s42 & p1, p4                          & f10,f23         \\ \hline
Processing Pipelines & s3, s8, s39, s10, s14, s16, s20                                                                                                                                                        & p2                              & -               \\ \hline
Data Lake            & s1,s5,s6,s14,s16,s19,s22,s23,s28,s29,s33,s35,s37,s40,s41,s42                                                                                                                           & -                               & -               \\ \hline
Event Store          & s31, s3, s4, s36, s14, s20, s21, s25, s27                                                                                                                                              & -                               & -               \\ \hline
Feature Store        & s19, s20, s22                                                                                                                                                                          & -                               & -               \\ \hline
\end{tabular}
}
\end{table*}

\subsection{ADD 3: Type of Data Ingestion}
\begin{figure*}[t]
    \centering
    \includegraphics[width=0.8\linewidth]{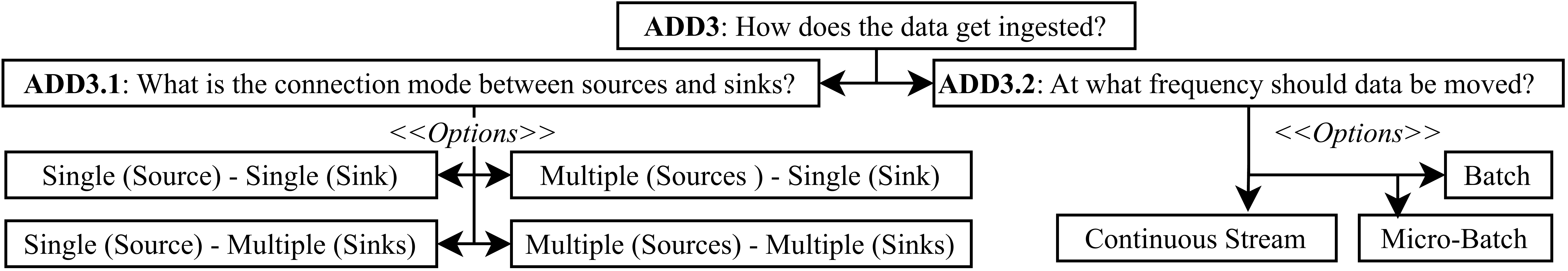}
    \caption{ADD 3: How does the data get ingested?}
    \label{fig:DI_ADD_3}
\end{figure*}

The third ADD considers how the data is ingested (see Figure~\ref{fig:DI_ADD_3} and Table~\ref{table:DI_ADD_3}). Two main decisions need to be made: \textbf{communication relationships between sources and sinks (ADD 3.1)}, and \textbf{data transmission rate (ADD 3.2)}.  In general, \textbf{\textit{m} data sources can be connected to \textit{n} data sinks (ADD 3.1)}. However, the literature mentioned specific cases that depend on the data architecture used. In the Lambda big data architecture~\cite{davoudian2020big}, data from the source is mirrored and routed to two parallel paths: a cold path for batch processing to provide historical insights and a hot path for real-time insights. In contrast, the Kappa architecture~\cite{davoudian2020big} has only a hot path, thereby forming a one-to-one relationship between the source and sink. However, in both architectures, data from multiple sources can be ingested simultaneously, e.g., by joining multiple data streams in Kappa or by extracting data from multiple sources in Lambda.

\begin{table*}[htbp]
\centering
\caption{ADD 3: options, evidence, and forces}
   \label{table:DI_ADD_3}
\resizebox{.99\textwidth}{!}{%
\begin{tabular}{|p{0.15\linewidth}|p{0.4\linewidth}|p{0.10\linewidth}|p{0.3\linewidth}|} 
\hline
\textbf{Options}          & \textbf{Source Evidences}                                                                                                                                                               & \textbf{Practitioner Evidences} & \textbf{Forces}                                                                                                                                \\ \hline
Single-Single            & s1, s2, s3, s4, s5, s6, s7, s8, s9, s10, s11, s12, s14, s15, s16, s17, s18, s19, s20, s21, s22, s23, s24, s25,s26, s27, s28, s29,$\sim$s30, s31, s32, s33, s34, s35, s36, s39, s40, s41 & p1, p5                          & f35(o), f6(-), f22(o), f7(o), f25(+), f15(+),f19(-), f2(o), f4(o), f14(+), f24(+), f30(+),f18(-), f5(o), f3(+), f28(+), f31(+), f34(+), f23(+) \\ \hline
Single-Multiple          & \multirow{2}{=}{s2, s4, s8, s12, s15, s16, s19, s31}                                                                                                                               & \multirow{2}{=}{p5}             & \multirow{2}{=}{f35(+), f7(+), f2(+), f4(+), f34(+), f23(+) }                                                                             \\ \cline{1-1}
Multiple-Multiple/Single &                                                                                                                                                                                         &         -                        &                                                                                                                                                \\ \hline
Batch                    & \multirow{2}{=}{s1, s2, s5, s8, s14, s18, s19, s20, s21, s22, s23, s25, s27, s30, s33, s36, s37, s39}                                                                                   & \multirow{2}{=}{-}              & f2(+), f10(+)                                                                                                                                  \\ \cline{1-1} \cline{4-4} 
Continuous Stream        &                                                                                                                                                                                         &                                 & f4(+), f30 (+)                                                                                                                                 \\ \hline
Micro-batch              & s13, s28, s30                                                                                                                                                                           & -                               & f5(+), f28(+)                                                                                                                                  \\ \hline
\end{tabular}
}
\end{table*}

Once the connection between the sources and destinations is established, the data should be transmitted. The literature discusses three \textbf{modes of data transmission (ADD 3.2)}: \textit{batch, micro-batch, and continuous stream}. The batch ingestion occurs when data are accumulated to a predefined size or over a fixed time interval, in which case all data within that interval are grouped into a batch. It is particularly useful when the data volume is large, but latency is not the primary concern. When low latency is required and newly generated data must be ingested into a destination immediately after generation, it must be ingested in real time as a stream. The amount of data ingested here can be relatively small, but the ingestion speed must be high and reliable. Connectivity is a primary concern for such real-time ingestion, as the data is usually ingested only once, with no backup, and at high speed, which means any connection fluctuation could cause data loss and degrade accuracy if no measures are taken to guarantee the connection and validate the data. The middle ground between batch and real-time ingestion is micro-batch ingestion. The data size here is smaller than that of batch ingestion but larger than that of real-time ingestion. This micro-batch approach is typically used in the data lakehouse architecture as an approximation (or alternative) to a continuous data stream processing model.

\subsection{ADD 4: Implementation of Data Ingestion}
\begin{figure*}[htbp]
    \centering
    \includegraphics[width=0.8\linewidth]{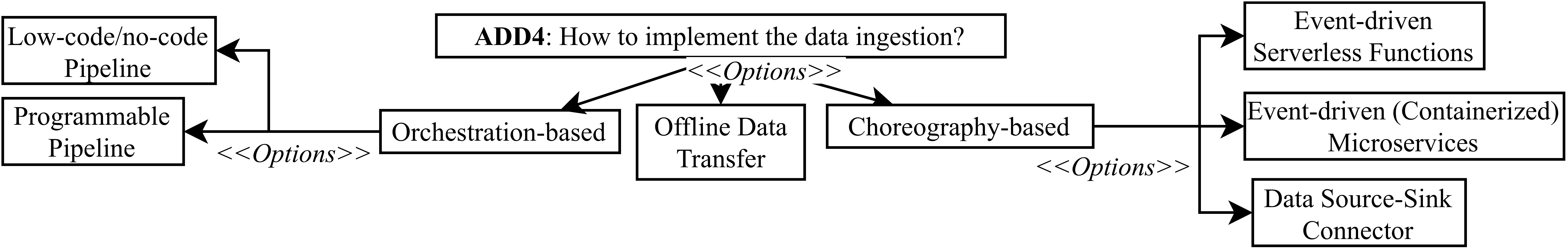}
    \caption{ADD 4: How to implement the data ingestion?}
    \label{fig:DI_ADD_4}
\end{figure*}

One of the crucial design decisions is how to implement the data ingestion in the overall data architecture. The reviewed gray literature discussed three options: \textit{orchestration-based implementation, choreography-based implementation, and offline data transfer} (see Figure~\ref{fig:DI_ADD_4} and Table~\ref{table:DI_ADD_4}).

An \textit{orchestration-based implementation} uses a workflow/pipeline definition and an orchestration tool to execute the workflow for data ingestion. The first option for implementing this workflow is to use a \textit{low-code/no-code approach} via drag-and-drop user interfaces (e.g., Azure Data Factory). This option is considered the most common and easiest-to-use implementation, allowing users to create, configure, and connect predefined components through a well-designed user interface. The workflow can also be implemented in a \textit{general-purpose or domain-specific language} provided by the tool provider (e.g., Dagster and Apache Airflow).  This provides greater flexibility to the users and enables organizations to manage security, performance, and quality concerns independently. 

\begin{table*}[htbp]
\caption{ADD 4: options, evidence, and forces}
\label{table:DI_ADD_4}
\scriptsize
\resizebox{.99\textwidth}{!}{%
\begin{tabular}{|p{0.17\linewidth}|p{0.40\linewidth}|p{0.10\linewidth}|p{0.30\linewidth}|} 
\hline
\textbf{Options}        & \textbf{Source Evidences}                                                                                                                                                                            & \textbf{Practitioner Evidences} & \textbf{Forces}                                                                                                                                                                        \\ \hline
1.Orchestration-based   & p1, p2, p3, p4, p5, s1, s2, s3, s4, s5, s6, s7, s8, s9, s10, s11, s12, s14, s15, s16, s17, s18, s19, s20, s21, s22, s23, s24, s25, s26, s29, s31, s32, s33, s34, s35, s37, s38, s39, s40, s42 & p1, p2, p3, p4, p5     & f35(-), f36(+), f29(o), f27(o), f26(+), f10(+), f12(o), f22(+), f7(+), f25(+), f19(o), f2(+), f16(+), f4(+), f14(o), f32(o), f18(-), f5(+), f9(+), f31(+), f34(+), f23(-), f11(+) \\ \hline
2.Choreography-based    & p1, p5, s3, s4, s6, s8, s9, s10, s11, s12, s13, s14, s16, s18, s19, s20, s21, s22, s24, s26, s27, s28, s29, s32, s33, s34, s36, s37, s40, s41, s42                                            & p1, p5                 & f35(+), f20(-), f36(-), f27(o), f10(+), f22(o), f7(o), f25(o), f15(o), f19(+), f2(o), f16(-), f4(-), f14(o), f24(+), f32(o),$\sim$f5(o), f3(-), f28(+), f34(o), f23(+)                 \\ \hline
3.Offline Data Transfer & s5,s9                                                                                                                                                                                         & -                      & f27(+), f6(+), f22(+), f2(+), f34(+), f23(+)                                                                                                                                           \\ \hline
\end{tabular}
}
\end{table*}

The \textit{choreography-based approach} primarily involves direct communication between data sources and destinations. Such communications are typically established using events and an event broker. There are three sub-options: \textit{data source-sink connector, event-driven microservice choreography, and event-driven serverless functions}. 

The first option is to \textit{use available plugins or connectors from external service providers or commercial/open-source platforms} to establish connections between data sources and sinks (e.g., Apache  Kafka connectors). This option requires compatibility between plugins or connectors and both the data source and the sink. The second option is a \textit{microservice architecture} for implementing event-driven data ingestion. This option facilitates the creation of customized, finely grained microservices that implement different steps in the data ingestion process (e.g, reading data, validating data, and uploading data). The third approach to implementing the choreography-based model is to \textit{use a chain of serverless functions}, each performing distinct ingestion logic. 

Compared to the choreography-based solution, the orchestration enables easy configuration of loosely coupled data sources, destinations, and other data platform components. The choreography-based implementation can give data engineers considerable control over the underlying ingestion solution. It is a solution characterized by its flexibility, reliability, and ease of configuration to meet exigent demands in security, privacy, parallelism, and concurrency. Nonetheless, developing such solutions typically requires expertise and can therefore be more complex and costly than orchestration. Additionally, the choreography-based approach may incur higher latency and introduce tight coupling between sources and destinations. 

The last option for implementing the data ingestion pipeline is the use of an \textit{offline data transfer device}, which is rarely mentioned in the gray literature because it is a limited offering from a small number of cloud providers. An example of data ingestion using this solution is migrating on-premises data storage to the cloud using dedicated hardware (e.g., AWS Snowball devices). The device is responsible for loading on-premises data and efficiently transferring it to cloud storage. It is particularly useful for ingesting large sensitive data volumes.

\subsection{ADD 5: Triggers of Data Ingestion} 
\begin{figure*}[t]
    \centering
    \includegraphics[width=0.9\linewidth]{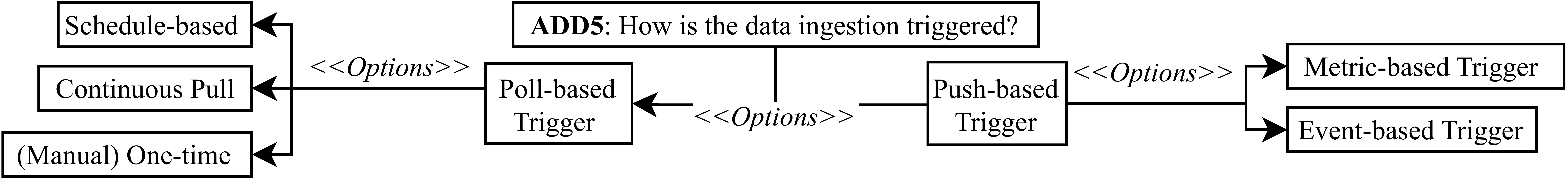}
    \caption{ADD 5: How is the data ingestion triggered?}
    \label{fig:DI_ADD_5}
\end{figure*}

Once a data ingestion pipeline is created, the next step is to decide how to trigger its execution. From the gray literature, we identified two options: \textit{poll-based and push-based triggers} (see Figure \ref{fig:DI_ADD_5} and Table \ref{table:DI_ADD_5}). Note that a data architecture may use one or both options. 

With the poll-based trigger, the data ingestion destination actively attempts to retrieve data from the ingestion source. Batch data ingestion typically \textit{uses a scheduling service} to pull data periodically or at a predefined time. The time interval is typically set to a day, week, or month, depending on data volume, usage, and other requirements. Stream data ingestion assumes a continuous stream of data from a source, then \textit{continuously pulls data} from the source, processes it, and pushes the processed data to data sinks (an unbounded ingestion cycle). The third variation is the \textit{manual on-time execution} of the ingestion pipeline. Common use cases include testing during the development phase of the data ingestion framework, or when data migrates from a source, such as an on-premises data center, to a destination, such as the cloud, requiring ingestion only once, with no further follow-up operations needed at that time. Overall, the poll-based trigger is effective for ingesting large volumes of data and enhances scalability and reliability in data ingestion solutions.

\begin{table}[t]
\centering
\caption{ADD 5: options, evidence, and forces}
\label{table:DI_ADD_5}
\resizebox{.99\linewidth}{!}{%
\begin{tabular}{|p{0.15\linewidth}|p{0.45\linewidth}|p{0.15\linewidth}|p{0.15\linewidth}|} 
\hline
\textbf{Solution}  & \textbf{Source Evidences}                                                                                                                 & \textbf{Practitioner Evidences} & \textbf{Forces}               \\ \hline
Poll-based Trigger & s1, s2, s5, s6, s7, s8, s10, s12, s15, s16, s17, s20, s21, s23, s24, s25, s26, s27, s29, s31, s32, s33, s34, s35, s36, s37, s38, s39, s41 & p5                              & f13(-), f2(+), f16(-), f34(+) \\ \hline
Push-based Trigger & s1, s2, s4, s5, s6, s7, s9, s11, s12, s14, s16, s17, s20, s24, s26, s27,s28, s34, s36, s39, s41                                           & p5                              & f27(+), f22(+), f13(+), f5(+) \\ \hline
\end{tabular}
}

\end{table}

With the push-based trigger, the data source initiates the ingestion process when data is generated. A common implementation option is to deploy and expose the ingestion pipeline as an endpoint, and then register it with the data source or an intermediary event broker.  The \textit{triggering can be based on generated events and monitored metrics}. For example, when an event broker receives generated events about newly produced data from a data source, it automatically triggers the subscribed data consumers, like an ingestion pipeline. A system that monitors an ML model's inference performance may trigger the ingestion pipeline to feed new data to retrain the underperforming model. While the push-based trigger processes data as it is generated at the source (enabling low latency and real-time insights), a lack of support for regulating the ingestion rate can overwhelm downstream consumers.

\subsection{ADD 6: Monitoring, Management and Governance of Data Ingestion} 
\begin{figure*}[t!]
    \centering
    \includegraphics[width=0.8\linewidth]{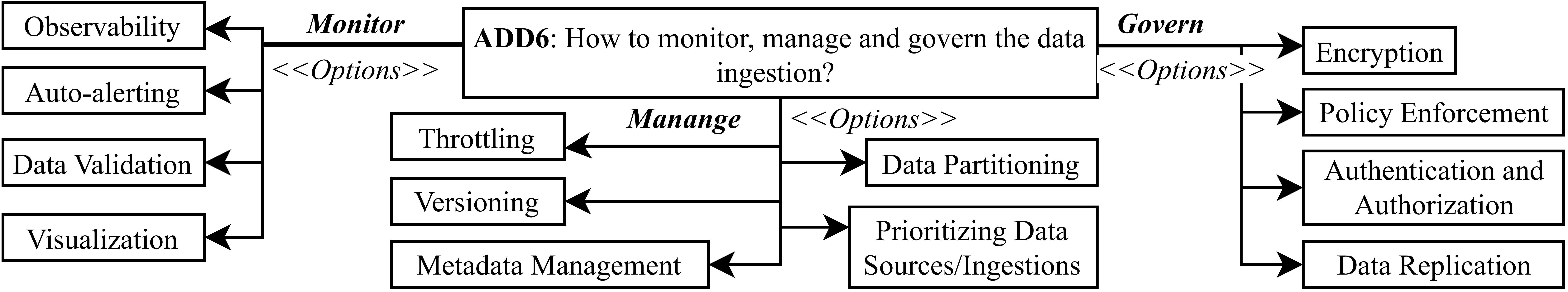}
    \caption{ADD 6: How to monitor, manage, and govern the data ingestion?}
    \label{fig:DI_ADD_6}
\end{figure*}

In addition to simply transferring data from source to destination, data ingestion processes within a robust big data/ML platform should be capable of performing ancillary tasks to meet the specific requirements of the platform's data ingestion users, including data security and quality. These requirements necessitate \textit{monitoring, management, and governance of data ingestion}. Figure~\ref{fig:DI_ADD_6} and Table~\ref{table:DI_ADD_6} show the tasks mentioned in the gray literature that aid in meeting these requirements.  

Regarding \textit{monitoring ingestion pipelines}, first, \textit{the logs, metrics, and traces} from the pipelines should be collected and analyzed to identify potential issues such as pipeline failures, anomalous behaviors (e.g, data volume spikes or data drops), and performance bottlenecks. Second, \textit{alerting} can inform the operational team about unexpected events and anomalies, including changes in pipeline, source, and destination states. Alerts can also trigger automated remediation processes, e.g., increasing resource allocation when the pipeline is overloaded or running a data transformation process when the data source schema changes.  Third, \textit{visualizing} the data ingestion process and the data itself can provide a clearer view of the data sources and sinks, their complex relationships, bottlenecks, and other issues, thereby improving the overall data ingestion environment. Finally, \textit{data validation} can help detect quality issues in data sources (e.g., missing or duplicate records) and ensure the correctness of data ingestion, for example, that the data was extracted completely (no data loss) and that the same record was not ingested twice (no data duplicates). The data validation can be implemented in the data source, during ingestion, or in the destination~\cite{ipekdataquality}.  

\begin{table*}[htbp]
\centering
\scriptsize
\caption{ADD 6: options, evidence, and forces}
\label{table:DI_ADD_6}

\begin{tabular}{|m{0.2\linewidth}|p{0.4\linewidth}|p{0.15\linewidth}|p{0.15\linewidth}|} 
\hline
\textbf{Options}                    & \textbf{Source Evidences}                                                                             & \textbf{Practitioner Evidences} & \textbf{Forces}             \\ \hline
Observability                       & s1, s4, s6, s7, s8, s11, s13, s14, s15, s18, s19, s20, s21,s22, s23,s24, s29, s36, s37, s39, s41, s42 & p1, p5                          & f15(+), f18(+)              \\ \hline
Auto-alerting                       & s29, s32, s11, s42, s13, s15, s17, s18, s22, s24                                                      & p5                              & f10(+),f11(+),f15(+),f29(+) \\ \hline
Visualization                       & s1, s3, s4, s6, s7, s8, s9, s10, s11, s13, s14, s15, s18, s20, s22, s23, s29, s31, s34, s35, s37      & -                               & f10 (+), f11(+)f21(+)       \\ \hline
Data Validation                     & s1, s31, s5, s6, s7, s8, s9, s10, s11, s12, s13, s14, s42, s15, s16, s32, s34, s36, s39               & p3, p5                          & f32(+),f20(+)               \\ \hline
Versioning                          & s19, s32, s40, s41, s42                                                                               & -                               & f32(+)                      \\ \hline
Metadata Management                 & s29, s31, s32, s34, s9, s39, s40, s41, s12, s15, s21, s23, s24, s28                                   & p2, p3, p4, p5                  & f1(+),f21(+)                \\ \hline
Data Partitioning                   & s34, s36, s40                                                                                         & -                               & f30(+),f35(+)               \\ \hline
Throttling                          & s13, s24                                                                                              & -                               & f9(+),f28(+),f35(+)         \\ \hline
Data Replication                    & s29, s32, s4, s5, s36, s13, s14, s15, s16, s18, s21                                                   & p5                              & f9(+),f18(+),f35(+)         \\ \hline
Prioritizing Data Sources/Ingestion & s29, s1, s5, s7, s8, s23                                                                              & -                               & f22(+),f30(+)               \\ \hline
Encryption                          & s32, s35, s5, s38, s39                                                                                & -                               & f23(+),f33(+)               \\ \hline
Authentication and Authorization    & s29, s32, s35, s39, s21, s23, s25, s27                                                                & p1, p2                          & f23(+),f33(+)               \\ \hline
Policy Enforcement                  & s29, s32, s35, s39, s21, s23, s25, s27                                                                & p1, p3                          & f23(+),f33(+),f1(+)         \\ \hline
\end{tabular}

\end{table*}

The gray literature discussed several strategies for effective management of data ingestion processes. The data (in sources and sinks) and data ingestion pipelines can be versioned to \textit{support handling schema changes}, improve data reproducibility, and enable graceful pipeline migrations. A related option is to \textit{capture and store metadata} of data assets and pipelines.   Incorporating metadata, such as data source and destination schemas, can ensure data consistency between the source and destination during ingestion. A common tool mentioned by practitioners was the \textit{data catalog} as the metadata repository, along with a metadata population pipeline (either a separate pipeline or steps within the data ingestion pipeline).  

Data ingestion on a big-data/ML platform is a complex process that requires a range of optimization tasks. \textit{Data partitioning} is the process of dividing incoming data batches or streams into partitions to optimize storage and performance during data ingestion. Another important task is prioritizing data sources/ingestion, which is crucial given the numerous data sources typically present in a large-scale data platform. This may involve \textit{prioritizing data sources} in a unified view for all downstream use cases, or prioritizing them separately for different use cases. Throttling is yet another task that can help to regulate data ingestion rate and prevent failures. With ML platforms handling vast data amounts for model training, throttling is necessary to manage the volume of data ingested. 

Replicating ingested data is occasionally necessary to create a backup at the destination, typically for implementing data recovery. \textit{Replicated data} may be kept in its raw form to maintain data lineage, or replicated in its ingested form to another storage location, particularly for data migration from on-premises databases to cloud storage.  

Regarding security requirements, \textit{authentication and authorization} measures should be implemented to ensure that only authorized users and applications can perform data ingestion, particularly when handling sensitive data in production. \textit{Encryption} can provide an additional layer of protection when ingesting sensitive or high-value data traveling over public networks, even when the connection between the data source and destination is secured. Finally, \textit{policy enforcement} can automatically apply governance rules to ensure regulatory compliance, data security, and data quality. For example, a policy can ensure automatic masking of sensitive data attributes or discard data sources below a given quality score threshold.

\smallskip
\subsection{ADD 7: Data Engineering Pipelines in Data Ingestion.} 
\begin{figure}[t]
    \centering
    \includegraphics[width=\linewidth]{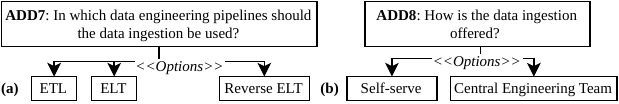}
    \caption{ADD 7 (data engineering pipeline) and ADD 8 (offering of data ingestion)}
    \label{fig:DI_ADD_7}
\end{figure}

The data ingestion process generally comprises three distinct operations, namely Extract (E), Transform (T), and Load (L). Subsequently, a crucial architectural decision concerns the number of these operations required for the data ingestion process and the order in which they must occur. While combining different numbers of these operations and altering their order can yield numerous solution options~\cite{reis2022fundamentals}, the most common and representative ones are \textit{Extract, Transform, and Load (ETL), Extract, Load, and Transform (ELT), and Reverse ETL} (see Figure~\ref{fig:DI_ADD_7}(a) and Table~\ref{table:DI_ADD_7}). 

\textit{ETL} is the conventional method for data ingestion that involves extracting data from its source, applying transformations, and loading the transformed data into a destination, typically a data warehouse (or its staging area). \textit{ELT}, on the other hand, is usually preferred by data engineers when data is ingested into a data lake in its raw form for ad-hoc data analysis pipelines. This method prioritizes loading data over transforming it, deferring the latter to a later stage. It reduces the complexity of ingestion solutions and increases flexibility, making it an increasingly popular option for big data and ML use cases. In contrast, \textit{reverse ETL} is a relatively new method used in specific use cases where processed data needs to be re-ingested from the final destination back to the sources. For example, user behavioral insights from the data warehouse are sent to the product ordering application to implement personalized recommendations.

\begin{table*}[t]
\centering
\caption{ADDs 7 and 8: options, evidence, and forces}
 \label{table:DI_ADD_7}
 \scriptsize
\resizebox{.99\textwidth}{!}{%
\begin{tabular}{|>{\centering\arraybackslash}m{0.03\linewidth}|m{0.12\linewidth}|p{0.35\linewidth}|m{0.10\linewidth}|p{0.25\linewidth}|} 
\hline
\textbf{ADD}       & \textbf{Options}         & \textbf{Source Evidences}                                                                                                                     & \textbf{Practitioner Evidences} & \textbf{Forces}                                                                            \\ \hline
\multirow{3}{*}{7} & ETL                      & s1, s2, s3, s4, s5, s7, s9, s10, s11, s12, s13, s14, s15, s16, s17, s18, s20, s21, s24, s25, s26, s27, s29, s31, s32, s34, s37, s38, s39, s41 & p1, p5                          & f35(+), f22(+), f25(o), f19(+), f16(+), f4(-),f32(+), f5(-), f34(+), f23(+), f1(+), f11(+) \\ \cline{2-5} 
                   & ELT                      & s1, s2, s3, s4, s7, s8, s9, s11, s12, s13, s15, s16, s17, s18, s19, s20, s21, s26, s27, s29, s31, s37, s38, s39, s40, s41                     & -                               & f35(+), f27(+), f22(+), f25(+), f15(+), f19(+),f4(+), f24(+), f32(o), f5(+)                \\ \cline{2-5} 
                   & Reverse ETL              & s26, s29                                                                                                                                      & -                               & f11(+)                                                                                     \\ \hline
\multirow{2}{*}{8} & Self-serve               & s17, s23, s40                                                                                                                                 & p2, p3, p5                      & f12(+), f36 (+)                                                                            \\ \cline{2-5} 
                   & Central Engineering Team & Most ETL and ELT sources imply a central platform team.                                                                                       & p1, p2, p3                      & f36 (-),f27 (+), f1(+)                                                                     \\ \hline
\end{tabular}
}
\end{table*}

\smallskip
\subsection{ADD 8: Offering of Data Ingestion.} When developing the data ingestion components for a data platform, it is important to consider how it will be offered to the potential platform users. Two possible approaches are a self-serve model, in which users have direct control over ingestion processes, or a centralized engineering team responsible for managing and maintaining the processes (see Figure \ref{fig:DI_ADD_7}(b) and Table~\ref{table:DI_ADD_7}). 

Providing a \textit{self-service} option for data ingestion enables a broad range of users, including data engineers, data scientists, and business owners, to independently configure, trigger, and monitor their data ingestion processes within the ML platform, without assistance or permission from dedicated data engineers. The development effort required to design and provision the data ingestion component in a self-serve manner may be greater, but it is expected to save significant time and cost for the data engineers in the operational phases of the deployed data platform. Alternatively, a \textit{centralized data engineering team} may offer ad hoc data ingestion within the data platform, configuring each use case individually to meet its specific technical and business requirements. This approach is commonly employed in early-stage data platforms and centralized data architectures, such as data lakes and data warehouses~\cite{goedegebuure2024data}. As the platform becomes mature, self-service features are incrementally added.   

\section{Expert Interview-based Validation and Refinement}
\label{sec:interview}

The subsequent sub-sections present the evaluation outcomes obtained from the expert interviews for the data ingestion architecture design decision framework. The sub-sections are structured as follows. Firstly, the experts' perceptions and considerations regarding the data ingestion component are discussed based on their professional expertise. This analysis aims to elicit the important and practically relevant aspects that experts would consider based on their work experience before being introduced to the framework in this study. Moreover, the experts' perceptions and considerations could serve as validation for the derived framework, its decisions, and various decision options~\cite{10592671}. The second subsection focuses on the framework's evaluation outcomes using the five evaluation criteria outlined in Section \ref{sec:research-design}: completeness, simplicity, understandability, operational feasibility, and usefulness. 

The feedback received from the expert interviews has been deemed valuable and insightful for enhancing the framework. The insights gathered through these interviews are expected to be leveraged for improving the quality of the framework, which in turn will enhance its efficacy and usefulness. The feedback from the experts has been carefully considered and analyzed to identify areas for improvement in the framework. The analysis of the feedback has been conducted with the objective of identifying common themes and patterns that can be used to refine and optimize the framework to meet the needs of its intended users more effectively. Overall, the feedback has been regarded as a valuable source of information to refine the framework and enhance its usefulness in practical settings.

\subsection{Perception and Considerations}

During the expert interviews, the interviewees were asked about their primary considerations when dealing with data ingestion in an ML platform for different use cases. The experts mentioned several common aspects, with all five emphasizing the source and destination of data ingestion, which the framework captures as the first and second architecture decisions. For instance, interviewee p2 emphasized that \textit{"...where is the data coming from depending on... That will affect a lot how the data can be ingested..."}, while p4 stated that \textit{"...it often involves some kind of landing zone where either a mechanism that you create yourself lands the data or where another team or company lands data for you..."}. These are clearly the first things practitioners, such as the interviewed experts, would consider when approaching a data ingestion use case. 

Depending on the organization the expert works in, they may emphasize different options regarding the source and destination. For example, p2, working in a manufacturing company in the Netherlands, considers the decision option of devices as data sources, which is the fourth option for the first decision in the data ingestion framework, more important than the others, whereas p1 mentioned that when working in a consultancy company and having a client from the finance sector, the source option of application which is the third option for the first decision in the data ingestion framework, gains the most attention. 

The third architectural decision, namely the type of data ingestion, was identified as another common architectural consideration by all five experts in machine learning engineering, as captured in the framework. The type of data ingestion, whether it is batch or stream ingestion or a more complex data architecture such as lambda architecture, significantly influences the implementation and the choice of technology solutions. Some experts have also highlighted the importance of selecting appropriate technology in practice, a point not fully addressed in the framework. For instance, expert p3 stated that \textit{"...it depends on the boundary condition that you have. For example, we have to operate in the context of some platforms that have been chosen for us in some cases, like Spark. So you don't have much of a choice in terms of which technology to use and how to ingest your data..."}. 

Interestingly, only one expert, p5, placed a substantial emphasis on quality and governance aspects of data ingestion, which is partly captured by the framework in the sixth architecture decision. The expert stated that \textit{"A big part of that is actually within the data ingestion area because there's a huge push for quality of data, not just in the data itself, but in the quality of the ingestion, so the quality of the pipeline. That's one part I think it's overlooked quite a lot, is that people focus on the data itself, what the data scientists would need, but they don't focus on the quality of the ingestion...They do not think about the quality of ingestion. So that's why I've started this new term, ingestion quality, just to basically say it's the quality of your pipeline."} The same expert advocates for the use of metadata to help ensure quality and govern ingestion, and also suggests that monitoring and validation mechanisms should be in place for ingestion jobs to improve the quality of the ingested data, which is the foundation and crucial for successful machine learning use cases.

\subsection{Framework Evaluation}
The present study reports the results of a qualitative evaluation of the data ingestion framework based on five distinct criteria, organized according to the communication structure proposed by \cite{shrestha2014communicate}. These authors emphasize the utility of using a matrix to analyze qualitative evaluation criteria. The feedback obtained from the semi-structured interviews was categorized as either positive, denoted by  A to indicate strong evidence of support for one of the evaluation criteria, or negative, represented by B to indicate strong negative comments regarding the evaluation criteria. 

\noindent Table \ref{tab:DI-EI-Results} provides an overview of the evaluation results for the five criteria discussed earlier, with respect to the data ingestion framework. The remainder of this sub-section provides a detailed discussion of each of the five criteria. 

\begin{table*}[h!]
    \centering
      \caption{DI Framework Evaluation}
        \label{tab:DI-EI-Results}
     % \resizebox{0.99\linewidth}{!}{
       \begin{tabular}{p{0.12\linewidth}|p{0.08\linewidth}|p{0.68\linewidth}} 
\hline
\textbf{Characteristic}                                           & \textbf{No. comments} & \textbf{Prominent comments} \\  \hline
Completeness                                                      & A x 12~ B x 4         & \textit{p5: "...actually I think is quite complete. I haven't seen too much that I'd want to add to that."p2: "there's something missing that you have to consider while doing the data ingestion..."}    \\ \hline
Simplicity                                                        & Ax 10 B x 4           & \textit{p2: Ï think it is simple enough to get started. Like for me, having already worked through }\textit{the process of implementing a data injection pipeline, it was relatively easy to go through it... "p1: "...but quite some options will not be relevant. "}  \\ \hline
Understandability                                                 & A x 15 B x 12         & \textit{p2: SSo the most of the text that is written in the..., so the text inside the boxes, I could understand."p3: "Data engineers should be okay. If you're talking about machine learning engineers, }\textit{I'm a bit worried. The terminology is different. "} \\ \hline
Operational feasibility & A x 12 B x 5          & \textit{p5: SSo this, I said, I would love to steal this and just start, every time someone}\textit{wants to do pipeline, I'm just gonna go through this and just start whacking up questions. "p4: Ï think the translation from a conceptual model to an actual implementation, }\textit{that is sometimes rather different because some implementations of a product }\textit{have other assumptions about how you think should do or think should work }\textit{compared to your conceptual model. "}  \\ \hline
Usefulness                                                        & A x 14 B x 4          & \textit{p1: Ï think it's useful. Certainly when you don't have any experience with data ingestion...}\textit{it also makes sense to check such a framework just to validate if they }\textit{have everything. If they have everything considered basically. "p4: Ï think when you have some kind of mechanism, so to say to}\textit{~solve this problem, a diagram doesn't actually solve the problem}\textit{~for the people who should use it. "}                                                       \\ \hline
\end{tabular}
      % }
        % \begin{tablenotes}
        %      \item \quad \done \quad  strongly supported by the expert in a comment 
        %      \item \null \quad \wontfix \quad  strongly opposed by the expert in a comment
        % \end{tablenotes}
         % \end{threeparttable}
    \end{table*}

\subsubsection{Completeness and Simplicity} 

Upon evaluating the data ingestion framework, experts in the field were asked to assess its completeness and simplicity. The results, as presented in Table \ref{tab:DI-EI-Results}, reveal overall positive feedback, with more positive comments on the framework's completeness and simplicity (12 and 10, respectively) than negative ones (4 and 4). The consensus among all experts is that the framework effectively captures the fundamental architectural design decisions and options applicable to various data ingestion use cases in an ML platform. Additionally, they agree that the framework presents these key factors in a simple, straightforward, and comprehensible manner. 

During the interviews, expert p3 stated that the data ingestion framework presented in this study is quite comprehensive and captures almost all available implementation options for a standard data ingestion use case. He commended the systematic approach taken to document all of the decisions and options, stating that he had a positive impression when reviewing the framework. Similarly, expert p5 agreed that the framework was complete and indicated that he had not seen much that he would want to add to it. He further commented that the questions presented in the framework were similar to those he asks people all the time, indicating its practical relevance as well. The same expert praised the individual diagrams within the framework, which allow practitioners to quickly and easily reference each decision and option. Overall, the experts provided positive feedback on the completeness and simplicity of the data ingestion framework. 

Expert p5 raised an important concern about the completeness of the data ingestion framework for streaming ingestion cases. While he appreciated the comprehensiveness of the framework, he believed that the seventh decision, which pertains to data engineering pipeline options, lacked an option specifically tailored for streaming ingestion cases: \textit{"So that's about the only thing I can really think of there that's missing ... Yeah, so there is ETL, but, you know, there's also a lot of cases now where the E in ETL, also the extract does not really happen so much in streaming data. ... There's not really like an ETL job running. So for streaming your space. It's streaming analytics, basically, data streaming pipelines or something."}. He pointed out that the existing options in the framework, including ETL, ELT, and reverse ETL, primarily cater to batch data ingestion scenarios, and an additional option for streaming use cases is imperative to enhance the framework's overall completeness. He argued that this was an important consideration, given the increasing prevalence of real-time data processing and analysis.

Expert p1 and p3 expressed their views on the simplicity level of the data ingestion framework in the context of real-world use cases. According to expert p1, although the framework captures a wide range of options, some of them may not be relevant to specific use cases. As an example, he cited a cloud-native team implementing a cloud-based data ingestion use case that would not require options that are not cloud-related. Therefore, the team may find the framework too complex to adopt. Expert p3 echoed similar concerns about the framework's simplicity, stating that \textit{"Not all of the use cases require so much. It might be confusing for somebody."}. He also pointed out that constraints might exist in some use cases where the freedom to choose basic options for building a solution may not be available, leading to negative comments about the framework's simplicity. Thus, while the framework may be comprehensive, it may not be universally applicable in all use cases. 

\subsubsection{Understandability}
The experts interviewed expressed mixed feelings when asked about the understandability of the data ingestion framework. Table \ref{tab:DI-EI-Results} reflects this, with 15 comments indicating positive understanding of the framework and 12 indicating negative ones. It is noteworthy that all interviewed experts explicitly stated that they found the framework's decisions and decision options easy to comprehend. However, it was observed that the Machine Learning Engineers (i.e., p2 and p3) made more negative comments regarding the framework's understandability characteristic than the other experts (i.e., Data Architect p1, Data Engineer p4 and p5). 

The interviewed experts generally agree that the data ingestion framework is easy to understand, especially for practitioners specialized in big data-related fields, namely experts p1, p4, and p5. They agree that the framework captures the high-level decisions and decision options, which should not pose any issues for practitioners to understand. A comment expert p5 made, which is quoted here: \textit{"...this is simple like (for example), you could give this to me when I just woke up in the morning and (I could) understand it."} shows that the framework is indeed straightforward for practitioners to understand, in his opinion. However, expert p3 and p4 noted the possibility of using different terms for certain decisions and decision options when practitioners work in different sectors or organizations, or even in different teams within the same company. Despite this, expert p4 also expressed confidence that his colleagues would comprehend the fundamental concepts underlying the framework even if slightly different terms are used for certain options. 

The experts who made negative comments emphasized that the framework might be difficult to understand for some practitioners who lack knowledge of data ingestion techniques or who have little experience in the field, like most of the data scientists, junior data engineers, or junior ML engineers. Expert p3 mentioned that "... The terminology is different. These concepts are not owned by data scientists, most of them... But as you know, you have teams where data scientists implement and ingest part of the data. And the knowledge about these concepts is quite deep. In my experience, I haven't seen many data scientists knowing these things...". He highlighted that some terms used in the framework could have different meanings across data science and data/ML engineering fields, which may lead to confusion and misalignment in a team. Expert p3 also noted that some data scientists may be unfamiliar with the concepts presented due to their specialization in other areas. Some experts, as also mentioned earlier, suggested that the framework's level of detail might be overwhelming, and some of the options presented might not be relevant to all use cases. 

It is essential to note that most experts agreed that practitioners with experience in big data-related fields should not have difficulty understanding the framework. Several experts even offered suggestions to improve the understandability of the data ingestion framework. Expert p2, p3, and p5 agreed that attaching brief explanations to the decisions and decision options could enhance the framework's understandability characteristic. Additionally, experts p2, p4, and p5 proposed that providing information on relevant technologies, existing market solutions, or specific use cases for each decision and each decision option would help practitioners better understand the framework. Such contextual information can provide a more practical understanding of the framework, particularly for individuals who are less familiar with data ingestion techniques or have limited experience in the field. 

\subsubsection{Operational Feasibility and Usefulness}

Experts in the field were also asked to assess the framework's operational feasibility and usefulness. The results, which are similar to the completeness and simplicity evaluation results, as presented in Table \ref{tab:DI-EI-Results}, reveal overall positive feedback, with more positive comments identified for operational feasibility and usefulness of the framework (12 and 14, respectively) compared to negative ones (5 and 4). This observation indicates that practitioners are inclined to adopt the proposed data ingestion framework. 

Expert p5 conveyed a highly favorable outlook regarding the operational feasibility of the data ingestion framework, as elucidated in his comment, \textit{"I would love to steal this and just start, every time someone wants to do (data ingestion) pipeline, I'm just gonna go through this and just start whacking up questions."} In his view, integrating the framework into the practical realm could facilitate data engineers in communicating more systematically and efficiently with other stakeholders, such as data scientists and ML engineers. The framework's decisions and decision options could guide data engineers to ascertain the pertinent questions to determine user requirements in data ingestion use cases. This could ultimately lead to a more structured and streamlined approach towards data ingestion. 

Expert p4 expressed doubts regarding the operational feasibility of the data ingestion framework derived in this study in an actual ingestion use case. However, he acknowledged its potential usefulness in educating junior data engineers who are new to the field. The expert reasoned that the transition from a conceptual model to an actual implementation can be challenging since some product implementations may have different assumptions about how the data ingestion solution should work compared to the conceptual model. Therefore, he was skeptical about the practicality of adopting the framework in a client project requiring a reliable and efficient data ingestion solution. He highlighted that experience holds greater significance in such situations than the framework.

Regarding the usefulness of the data ingestion framework, experts p1 and p4 opined that it would be more valuable for novice data and ML engineers for both educational and validation purposes than for senior ones. Expert p1 remarked that \textit{"I think it's useful, especially when you have no experience with data ingestion...it also makes sense to use such a framework to validate if all relevant aspects have been considered."} They concurred that the framework could serve as a comprehensive checklist at the beginning of a data ingestion use case for documenting all pertinent user requirements or at the end for validating the developed solution against the framework. However, expert p3 expressed apprehension that junior engineers might not comprehend all the framework's contents, thereby impairing its usefulness in requirement documentation and validation scenarios. 

Experts p1 and p2 highlighted that changes in requirements are frequent occurrences during a data ingestion use case, and they believe that the framework's usefulness in addressing requirement changes is limited. They noted that the framework does not offer a mechanism for adjusting the data ingestion architecture from the initial decision options to updated decision options that meet the new requirements. They opined that repeating the entire framework evaluation in a use case each time a requirement change occurs is burdensome and inefficient. Thus, they suggested that if the framework could offer an approach for managing requirements and decision option changes, it would enhance its usefulness substantially.

\section{Threats to Validity}
\label{sec:discussion}

% This study is not immune to the common threats to validity in qualitative research.
This section discusses the potential threats and mitigation strategies in terms of internal, construct, and external validity.

\smallskip
\noindent\textbf{Internal Validity.}
The main threats to internal validity involve potential biases in the literature analysis and interview-based evaluation. In the gray literature review, source relevance and quality could affect the findings' reliability. To mitigate this, we applied quality assessment criteria, examined references and backlinks, and conducted independent source selection by two authors, calculating inter-rater agreement and retaining only sources with consensus.

Coding biases and interpretation errors pose another threat. We mitigated these by following established qualitative analysis guidelines, including saturation criteria. The first author coded all articles, while the second reviewed the codes in several meetings, resolving disagreements through consensus.

The interview-based evaluation introduced potential threats from participant selection bias, expertise relevance, and data analysis inconsistencies. To mitigate these, we selected participants from diverse stakeholder groups and verified their professional backgrounds. Despite the limited sample size, their varied expertise provided relevant perspectives for evaluating and complementing the literature findings.

% \smallskip
\noindent\textbf{Construct Validity.}
Construct validity concerns whether the study's methods adequately capture the concepts under investigation. Potential threats may arise from the design of the interview questions and the interpretation of participants' responses. To address these issues, we developed a predefined interview protocol, and the questions were grounded in insights obtained from the literature review. Furthermore, the interview data were analyzed using a systematic thematic analysis approach to support consistent interpretation and improve the reliability of the findings.

% \smallskip
\noindent\textbf{External Validity}.
The main threat to external validity concerns the generalizability of the findings beyond big data pipelines. To maintain a feasible scope, we focused on data and machine learning pipelines, excluding emerging AI systems, such as generative AI architectures, which may introduce additional design considerations. Moreover, as the gray literature search was conducted in 2023, newer studies may reveal additional or updated ADDs, options, and drivers. Nevertheless, the synthesized knowledge, refined through practitioners' feedback, provides a substantial foundation of 8 ADDs, 59 options, and 36 drivers.

% Consequently, further research is required to assess whether the identified architectural decisions, options, and drivers apply to other types of data-intensive systems.

% A potential external threat is the generalizability of our findings beyond big data pipelines.  To reduce manual effort in analyzing the articles, we intentionally focused on sources related to data/ML pipelines and excluded emerging AI systems, such as generative AI-based architectures, where data ingestion may involve many design issues.   

\section{Conclusion and Future Work}
\label{sec:conc}
Data ingestion is a key element of a big data architecture, collecting data from various sources and moving it to downstream data processing and analytics layers. Data ingestion deficiencies can result in incomplete and inaccurate analytical data, leading to erroneous business insights and decisions.  In this paper, we presented an architectural design (ADD) framework to guide practitioners in systematically designing data ingestion processes in big-data architectures. Our framework includes 8 ADDs, 59 decision options, and 36 selection drivers. We created it through a systematic review and analysis of the gray literature on data ingestion, and we evaluated the framework with semi-structured expert interviews.  Future research will extend the framework with new sources and address data ingestion challenges in generative AI architectures. % and to build a tool to automatically update decision models by extracting the relevant information from the literature using generative AI.

% \section*{Data Availability Statement}
% \label{sec:dataavailability}
% The replication package for our study is available online\footnote{\url{https://doi.org/10.5281/zenodo.21510046}}. It comprises the complete list of sources, the criteria used to select and assess their quality, and the coded literature. It also includes the semi-structured interview protocol, their anonymized transcripts, and the corresponding interview coding, which support both the development of the framework and the evaluation of its utility.

%\section*{Acknowledgments}

%\onecolumn

\noindent\textbf{Acknowledgments:} This research has been partially funded by the RAPTURE TKI research project (RT104659-11)

\section*{CRediT Author Statement}

\noindent\textbf{Yikang Huang}: conceptualization, methodology, investigation, software, formal analysis, data curation, visualization, writing – original draft, writing – review \& editing. \textbf{Stefano Fossati}: validation, formal analysis, visualization, writing – original draft, writing – review \& editing. \textbf{Indika Kumara}: methodology, validation, formal analysis, data curation, visualization, writing – original draft, writing – review \& editing, supervision. \textbf{Damian A. Tamburri}: funding acquisition, supervision, writing – review \& editing. \textbf{Willem-Jan Van Den Heuvel}: funding acquisition, supervision, writing – review \& editing.

\bibliographystyle{IEEEtran}
\bibliography{main}
\vspace{-33pt}
\begin{IEEEbiographynophoto}{Yikang Huang} holds an MSc in Data Science from JADS.
%is a Ph.D. candidate at the Eindhoven University of Technology - JADS. His research focuses on software infrastructure, its observability, maintainability, and resilience. % He holds an MSc in Computer Science and Engineering. % Contact: \href{mailto:s.fossati@tue.nl}{s.fossati@tue.nl}.
\end{IEEEbiographynophoto}
\vspace{-33pt}
\begin{IEEEbiographynophoto}{Stefano Fossati}
is a Ph.D. candidate at the Eindhoven University of Technology - JADS. His research focuses on software infrastructure, its observability, maintainability, and resilience. % He holds an MSc in Computer Science and Engineering. % Contact: \href{mailto:s.fossati@tue.nl}{s.fossati@tue.nl}.
\end{IEEEbiographynophoto}
\vspace{-33pt}
\begin{IEEEbiographynophoto}{Indika Kumara}
is an assistant professor at JADS and Tilburg University, with research interests in data architecture, service-oriented computing.% and software architecture. %He holds a Ph.D. in Computer Science. 
\end{IEEEbiographynophoto}
\vspace{-33pt}
\begin{IEEEbiographynophoto}{Damian Andrew Tamburri} is an associate professor at the University of Sannio and JADS, with research interests in DevOps architectures and software engineering. % He holds a Ph.D. in Computer Science. %Contact: \href{mailto:d.a.tamburri@tue.nl}{d.a.tamburri@tue.nl}.
\end{IEEEbiographynophoto}
\vspace{-33pt}
\begin{IEEEbiographynophoto}{Willem-Jan van den Heuvel}
is a full professor at JADS and Tilburg University. His research interests include data ecosystems, service-oriented computing, and software architecture.% He holds an MSc in Computer Science and Engineering. % Contact: \href{mailto:s.fossati@tue.nl}{s.fossati@tue.nl}.
\end{IEEEbiographynophoto}

\vfill

\end{document}